\documentstyle[epsf]{mn2e}
\begin{document}

\title{Semi-Regular Variables in the Solar Neighbourhood}

\author[I.S. Glass \& F. van Leeuwen]{I.S. Glass$^1$ \& F. van
Leeuwen$^2$ \\
$^1$South African Astronomical Observatory, PO Box 9, Observatory 7935, South
Africa\\
$^2$Institute of Astronomy, University of Cambridge, Madingley Road,
Cambridge CB3 0HA}

\date{Accepted 2007 April 20. Received 2007 April 18; in original form 2007
March 29}

\maketitle

\begin{abstract}

Period-luminosity sequences have been shown to exist among the Semi-Regular
Variables (SRVs) in the Magellanic Clouds (Wood et al, 1999), the Bulge of
the Milky Way galaxy (Glass \& Schultheis, 2003) and elsewhere. Using modern
period and revised Hipparcos parallax data, this paper demonstrates that
they also appear among the M-giant SRVs of the Solar Neighbourhood. Their
distribution in the $K$, log $P$ diagram resembles that of Bulge stars more
closely than those in the Magellanic Clouds.  The prevalence of mass-loss
among local M-type SRVs and its dependence on period and spectral sub-type
are also discussed. $K$ -- [12], a measure of circumstellar dust emission,
increases clearly with $V$ amplitude, M giant sub-type and log $P$.

\end{abstract}


\section{Introduction}

The large-scale surveys for Massive Compact Halo Objects (MACHO), the
Optical Gravitational Lensing Experiment (OGLE) and similar unbiased
searches have revolutionized our knowledge of the variable star populations
in the Galactic Bulge and the Magellanic Clouds. Previous work relied on
photographic surveys which usually revealed only objects having amplitudes
of several tenths of a magnitude or more. Thus, while our knowledge of Miras
was fairly complete in certain fields, this was not true of the SRVs, whose
amplitudes are usually much smaller. Small-amplitude SRVs have turned out to
be extremely numerous relative to large-amplitude ones whenever searches
have been sensitive enough to find them (e.g., Wood et al, 1999, for the
Large Magellanic Cloud; Alard et al, 2001, for the Bulge).

Although the period-luminosity relation for Mira variables has been known
for a long time, the analysis of MACHO data by Wood et al (1999) showed that
similar trends exist among the SRVs of the Large Magellanic Cloud. Wood
(2000) found that there are at least five clear sequences (A,B,C,D,E) in the
the $K$, log$P$ diagram. More refined observations by Kiss \& Bedding (2003)
and Ita et al (2004) have since made it clear that the A and B sequences
undergo perceptible `jogs' at about the level of the Red Giant Branch (RGB)
tip, requiring that they be sub-divided into A$^+$, A$^-$, B$^+$ and B$^-$.
In addition they found a sequence designated by the latter as C$'$ with the
suggestion that it is populated by first overtone Mira-like variables. The
near-infrared yields tighter $M$, log$P$ relations than the visible region
because of the diminished effect of interstellar reddening and the fact that
the amplitude of variation is less.

Glass \& Schultheis (2003) showed that SRV sequences are not confined to the
Magellanic Clouds but also occur in the NGC\,6522 Baade's Window field in
our own galaxy, though here they are smeared out due to the depth of the
Bulge. Lebzelter et al (2005) noted that a small number of nearby luminous
SRVs with Hipparcos parallaxes fall on the Wood B and C (Mira) sequences, as
do several SRVs in the globular cluster 47 Tuc. Recently, the first results
of a comprehensive survey of SRVs in globular clusters have been presented
by Matsunaga et al (2006). From their work, it is evident that the globular
cluster SRVs do not extend to such high luminosities as those in the
Magellanic Clouds.

Whitelock (1986) found a relation between $M_{\rm bol}$ and log$P$ for the
SRVs in the globular clusters 47 Tuc and NGC\,5927. Its slope agreed with
the evolutionary tracks of Vassiliadis \& Wood (1993). Similarly, in an
analysis of the $M_K$, log$P$ diagram of nearby SRVs, Bedding \& Zijlstra
(1998) suggested that they fit a line parallel to Whitelock's one but
$\sim$0.8 mag brighter. The slope they found is much shallower than those of
the $K$, log$P$ relations determined for LMC O-type Miras and for SRVs by Wood
(2000). It may, however, represent an evolutionary track for the most
luminous SRVs of the Solar Neighbourhood, though Glass \& Schultheis (2003)
suggested that it was probably an artefact arising from the poor knowledge
of SRVs with small amplitudes then prevailing.

Population-dependent trends have been searched for by Schultheis, Glass \&
Cioni (2004) by analysing MACHO data from the two Magellanic Clouds and the
NGC\,6522 field in a similar manner. They found that, as metallicity
decreases, the luminosity of the RGB tip decreases, the proportion of
variable stars decreases and the minimum period associated with a given
amplitude gets longer; i.e., the amplitudes are lower when the metallicity
is lower. The differences between the LMC and the SMC have recently been
investigated more thoroughly by Kiss and Lah (2006). Further, it is evident
that there are few luminous stars on the A$^+$ and B$^+$ sequences of the
Milky Way field, in part due to the absence of carbon stars.

In this paper the $M_K$, log$P$ diagram for nearby SRVs is re-examined in
the light of new information that has become available. A small number of
low-amplitude red variables have been discovered among the nearby stars and
monitored photoelectrically over substantial time intervals, with the result
that accurate periods have been determined for them. The Hipparcos parallax
data have also been re-reduced using improved methods so that the probable
errors in the distances of the red stars it observed are now approximately
halved.

\section{The present sample}

The sample of SRVs in this paper is drawn mainly from Percy, Wilson \& Henry
(2001), Percy, Dunlop, Kassim \& Thompson (2001), Percy, Nyssa \& Henry
(2001) and Percy et al (2004), in which periods and amplitudes are presented
for a number of SRVs derived from long series of data taken by themselves
and members of the American Association of Variable Star Observers. In
addition, some stars listed by Bedding \& Zijlstra (1998), Hron, Aringer \&
Kerschbaum (1997) and Olofsson et al (2002) have been included (note: g Her
= 30 Her) even though there may be no new information about their periods.

Table \ref{sample} shows the sample. As far as possible, it has been limited
to M stars of luminosity class III (as listed by CDS, Strasbourg) and
regarded as having SRa or SRb variability type. The individual stars in many
cases have extensive literatures of their own. To be useful for the present
purpose, each star must have a well-determined parallax, a $K$ magnitude and
good period information. As a result of the first criterion, they will be
nearby and therefore too bright to be present in the most recent near-IR sky
surveys, DENIS and 2MASS.

The source of the period information is given in the `Ref' column of the
table. Inevitably for SRVs, the determination of characteristic periods is a
lengthy affair, requiring observations stretching over many cycles (often
also involving long secondary periods). There is room for argument about the
correctness of the periods in many cases. The choice of periods to include
in the analysis is explained in the footnotes of Table \ref{sample}. This
follows as far as possible the approach used in previous work by one of the
the authors and his collaborators concerning SRVs in the Magellanic Clouds
and the NGC\,6522 field, in the sense that only one each of the predominant
short and long periods are retained.

\begin{figure}
\epsfxsize=8.5cm
\epsffile[28 285 539 786]{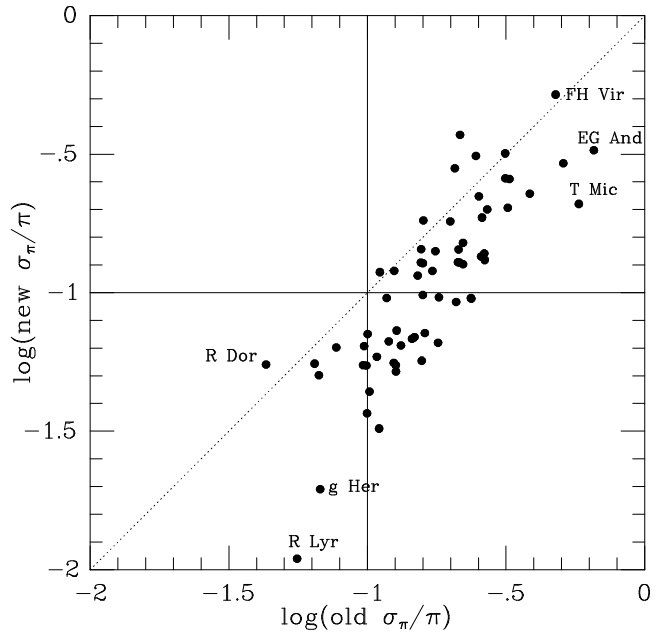}

\caption{Logarithm of the relative errors of the objects in Table 2
according to the old and new Hipparcos reductions. Objects
with ${\pi / \sigma_{\pi}} > 10$ are now much more numerous.
Some outlying points have been labelled.}

\label{floor}
\end{figure}

The distances of these stars have been derived from the revised Hipparcos
Catalogue (van Leeuwen \& Fantino, 2005; van Leeuwen, 2007). In this new
version, the errors for red stars have usually been reduced by about a
factor of two from the previously published Catalogue (ESA, 1997).
Fig \ref{floor} shows the improvements due to the new parallax reductions
(van Leeuwen \& Fantino, 2005; van Leeuwen, 2007).

$K$-band magnitudes have been taken, in general, from the {\it The
Two-Micron Sky Survey} of Neugebauer \& Leighton (1969). Each source was
observed a few times and the probable error of the magnitude was usually
about 0.04 mag. Some additional magnitudes were supplied by Dr T. Lloyd
Evans (University of St Andrews).

The amplitudes are from $V$ data, where available. Otherwise, $B$ or
photographic ones have been used.

\subsection{Biases in the sample}

The sample cannot pretend to be complete. Only a very few of the nearby
late-type SRVs have been observed photometrically with the necessary
precision and for sufficiently long times to determine their variability
characteristics.

Since the sample is selected to some extent on parallax error, the
Lutz-Kelker bias should be examined. The calculations of Koen (1992, case
$p$ = 2) may be used to determine the mean value of $\Delta M$, the
error in the estimate of the distance moduli. The mean value of
${\sigma_{\pi} / \pi}$ for the $\pi$ $>$ 3$\sigma_\pi$ sample (see
below) is $\sim 0.1$; according to Koen (Table 2), 0.07 mag is the average
amount by which stars will have been shifted downwards in Fig 1. The mean
for the $\pi$ $>$ 10$\sigma_\pi$ sample is $\sim$ .063; the average downward
shift in Fig 2 is then about 0.025 mag.

Glass \& Schultheis (2002), using observations made during the MACHO
gravitational lensing project, showed that essentially all M giants later
than MK subtype 4--5 in the Bulge NGC\,6522 field drawn from the complete
sample observed by Blanco (1986) are variable. Though there are certainly
differences in the M giant populations locally and in the Bulge, especially
considering the absence of carbon stars among the latter, it is reasonable
to expect that local stars of similar spectral types will be seen to vary.
The present sample is heavily biased towards the later M sub-types, as can
be seen from Table \ref{bsc}, which shows the numbers of each M giant
sub-class present in the Bright Star Catalogue (Hoffleit and Warren, 1991,
as found on-line at CDS, Strasbourg) and the numbers of these present in
Table \ref{sample}. (Intermediate sub-types such as M5.5III were not
considered.)

In making comparisons with the Magellanic Cloud and NGC\,6522 results, it
must be remembered that the IRC $K$ magnitudes use the traditional broadband
$K$ and not the $K_S$ band of DENIS and 2MASS. The latter filter does not
include the CO first overtone band which is prominent in late-type stars.
The absence of the 2.3$\mu$m absortpion it causes may make the local sample
appear a few hundredths of a mag fainter than if $K_S$ were used.

Further, no compensation has been applied for interstellar reddening to the
local $K$ values. It is estimated that this will not be more than one or two
hundredths of a magnitude in most cases.

\begin{table}

\caption{M giants in the Bright Star Catalog and number of these in the
present sample}

\begin{tabular}{lll}
Sub-type$^1$ & No.\ in BSC & of which, no.\ in Table \ref{sample} \\
M0 & 64 & 2    \\
M1 & 80 & 0    \\
M2 & 73 & 2    \\
M3 & 52 & 6    \\
M4 & 44 & 5    \\
M5 & 17 & 4    \\
M6 & 7  & 4$^2$\\
M7 & 0  & 0    \\
\end{tabular}

$^1$Note: sub-class M{\it n} includes M{\it n}III, M{\it n}IIIa, M{\it n}IIIb, 
M{\it n}IIIab.  

$^2$g Her is given as M6-III in the BSC.

\label{bsc}
\end{table}

\begin{table*}
\caption{Data for local M-type SRVs in sample}
\begin{tabular}{lllllllll}
\\
HD     & Name      & Sp       &	 Periods (d) & $\pi$ & $\sigma_{\pi}$ & $K$ &
Ampl.\ & Ref.\\
\
002411 & TV Psc    & M3III    &  55*                    &  6.16 & 0.59 & -0.16 & 0.5   & PWH     \\
003346 & V428 And  & M0III    &  11, {\bf 11.5}, 15, 22 &  5.28 & 0.30 & 1.20  & 0.065 & PNH     \\
004174 & EG And	   & M2e sym  &  29.1$^1$, 47.6         &  1.96 & 0.64 & 2.74  & 0.27  & P et al \\	
004408 & NSV 00293 & M4III    &  12?, 32?, 40?          &  4.20 & 0.29 & 0.15  & 0.22  & PDKT    \\
005820 & WW Psc	   & M2III    &  {\bf 25}, 300          &  3.03 & 0.42 & 1.58  & 0.23  & PDKT    \\
013596 & CSV 100168& M0III    &  32:, 275:              &  6.81 & 0.38 & 1.80  & 0.14  & PKDT    \\
017491 & Z Eri     & M5III    &  80                     &  3.87 & 0.49 & 0.32  & 1.63  & GCVS    \\
018191 & RZ Ari	   & M6III    &  37.7, 56.5$^1$	        &  9.28 & 0.30 & -1.08 & 0.4   & P et al \\
022689 & SS Cep	   & M5III    &  90	                &  3.84 & 0.49 & -0.56 & 1.1   & GCVS    \\
029712 & R Dor	   & M8IIIe   &  338	                & 18.35 & 1.01 & -3.91 & 1.8   & GCVS    \\
030959 & 4 Ori	   & M3 (S)   &  36$^1$, 52.6, 74.1     &  5.02 & 0.72 & -0.53 & 0.3   & P et al \\
033664 & RX Lep	   & M6III    &  60:/80,long	        &  6.70 & 0.43 & -1.25 & /0.5  &GCVS/PWH \\
039983 & BQ Ori    & M5III    &  110                    &  4.74 & 1.22 & 0.84  & 2.1   & GCVS    \\
041698 & S Lep     & M5III    &  89                     &  4.90 & 0.63 & -0.49 & 1.58  & GCVS    \\
042973 & UW Lyn	   & M3III    &  26.0, 37.6$^1$, 49.5$^1$& 5.12 & 0.33 & 0.73  & 0.15  & P et al \\
042995 & $\eta$ Gem & M3III   &  234*, shorter?         &  8.52 & 1.22 & -1.49 & 0.3   & PWH     \\
044478 & $\mu$ Gem & M3III    &  20, 27.0$^1$, 51.0     & 14.10 & 0.71 & -1.89 & 0.23  & P et al \\
051725 & V523 Mon  & M5	      &  26.0, 34.1$^1$, 45.6   &  2.83 & 0.63 & 1.34  & 0.2   & P et al \\
056096 & L$_2$ Pup & M5IIIe   &  140.6	                & 15.61 & 0.99 & -2.15 & 1.2   & BZ      \\
062647 & NSV 03721 & M3III    &  22:, 360               &  7.51 & 0.41 & 0.89  & 0.13  & PDKT    \\
064052 & BC CMi	   & M4III    &  20:, 28:$^1$, 45:      &  6.44 & 0.47 & 0.86  & 0.5   & P et al \\
073844 & AK Hya	   & M6III    &  50::	                &  6.37 & 0.42 & -0.57 & 1.16  & PDKT    \\
075716 & BO Cnc	   & M3III    &  {\bf 27}, 270          &  3.70 & 0.75 & 1.39  & 0.26  & PDKT    \\
077443 & UX Lyn	   & M3	      &  37.2$^1$, 51.3         &  4.16 & 0.63 & 0.48  & 0.4   & P et al \\
094705 & VY Leo	   & M5III    &  48, 500                &  8.42 & 0.37 & -0.80 & 0.75  & PDKT    \\
099592 & ST UMa    & M4/5III  &  50, 81*, 625*          &  1.38 & 0.43 & 0.58  & 0.7   & PWH     \\
101153 & $\omega$ Vir & M4III &  {\bf 30}, 275          &  6.57 & 0.36 & -0.27 & 0.28  & PDKT    \\
102159 & TV UMa	   & M4III    &  600	                &  4.34 & 0.81 & 0.83  & 0.72  & PDKT    \\
112264 & TU CVn    & M5III    &  44.5*, 230:            &  4.69 & 0.32 & -0.13 & 0.35  & PWH     \\
113285 & RT Vir    & M8III    &  155                    &  7.46 & 0.86 & -0.97 & 1.29  & GCVS    \\
113866 & FS Com	   & M5III    &  38.2, 55.4$^1$         &  4.43 & 0.41 & -0.21 & 0.35  & P et al \\
114961 & SW Vir	   & M7III    &  {\bf 155}	        &  7.01 & 0.84 & -1.74&1.85/1.8& PDKT/PWH\\
115322 & FH Vir	   & M6III    &  72, 280                &  1.33 & 0.69 & 1.45  & 1.19  & PDKT    \\
118767 & V744 Cen  & M5III    &  90	                &  6.35 & 0.33 & -0.75 & 1.41  & GCVS    \\
120285 & W Hya	   & M7e      &  361	                &  9.77 & 1.17 & -3.17 & 3.1   & GCVS    \\
122250 & $\theta$ Aps & M6.5III: & 119                  &  8.84 & 0.49 & -1.92 & 2.2   & GCVS    \\
124304 & EV Vir	   & M3III    &  {\bf 19.5, 57}         &  1.98 & 0.58 & 1.52  & 0.52  & PDKT    \\
124681 & FS Vir	   & M4III    &  20, 250                &  4.04 & 0.52 & 1.54  & 0.18  & PDKT    \\
125180 & CY Boo	   & M3III    &  {\bf 23}, 350          &  4.28 & 0.41 & 1.49  & 0.10  & PDKT    \\
126327 & RX Boo	   & M7.5     &  340	                &  4.98 & 0.64 & -1.85 & 2.7   & GCVS    \\
140297 & RR CrB	   & M3	      &  60.8	                &  2.93 & 0.53 & 0.94  & 1.7   & GCVS    \\
143347 & RS CrB    & M7       &  333                    &  3.05 & 0.40 & 1.77  & 2.9   & GCVS    \\
144205 & X Her	   & M8	      &  95.0	                &  7.31 & 0.40 & -1.48 & 1.1   & GCVS    \\
148783 & g Her	   & M6III    &  93*, 833*              &  9.22 & 0.18 & -1.99 & 0.6   & PWH     \\
150077 & TX Dra	   & M5	      &  78	                &  2.91 & 0.53 & 1.43  & 2.3   & GCVS    \\
151187 & S Dra     & M6III    &  136                    &  2.42 & 0.77 & 0.06  & 1.0   & GCVS    \\
151481 & AZ Dra	   & M2III    &  {\bf 352}              &  2.74 & 0.37 & 2.40  & 0.55  & JDKT    \\
152152 & AH Dra    & M5       &  158                    &  2.56 & 0.72 & 0.59  & 0.8   & GVCS    \\
159354 & V642 Her  & M4III    &  25.6, 35.7$^1$         &  5.41 & 0.52 & 0.93  & 0.29  & P et al \\
167006 & V669 Her  & M3III    &  {\bf 27}               &  5.99 & 0.22 & 0.35  & 0.17  & JDKT    \\
175865 & R Lyr	   & M5III    &  45.9, 64.1$^1$         & 10.96 & 0.12 & -2.10 & 0.6   & P et al \\
184008 & AF Cyg    & M4       &  92.5	                &  4.53 & 0.64 & 0.29  & 2.0   & GCVS    \\
184313 & V450 Aql  & M5/5.5III&  65*                    &  4.94 & 0.47 & 0.14  & 0.35  & PWH     \\
186776 & V973 Cyg  & M3III    &  {\bf 35, 376}          &  3.98 & 0.39 & 1.49  & 0.40  & JDKT    \\
194676 & T Mic     & M7III    &  347                    &  4.83 & 1.01 & -1.56 & 1.9   & GCVS    \\
195351 & UU Dra    & M8IIIe:  &  120                    &  2.99 & 0.68 & 0.38  & 1.5   & GCVS    \\
196610 & EU Del	   & M6III    &  59.7/62.3, long	&  8.53 & 0.50 & -1.10 &0.7/0.7&GCVS/PWH \\
201298 &           & M0III    &  12, 13, 40:            &  2.80 & 0.56 & 1.06  & 0.041 & PNH     \\
203712 & V1070 Cyg & M7III    &  110, 470/60, 50$\pm$, complex&6.21&0.44&-0.66 &0.83   &PDKT/PWH \\
205730 & W Cyg	   & M4III    &  130.4*, complex/131.1  &  5.70 & 0.38 & -1.35 & 1.0   &PWH/GCVS \\
207076 & EP Aqr    & M8IIIv   &  55                     &  8.82 & 0.63 & -1.55 & 0.45  & GCVS    \\
209872 & SV Peg    & M7       &  145                    &  4.20 & 1.56 & -0.55 & 1.8   & GCVS    \\
209958 & TW Peg    & M7.5IIIv &  929                    &  7.50 & 0.89 & -0.63 & 0.9   & GCVS    \\
215162 & BD Peg    & M8       &  78                     &  4.29 & 1.11 &  1.10 & 0.9   & GCVS    \\
\\
\end{tabular}
\end{table*}

\setcounter{table}{1}
\begin{table*}
\caption{Data for local M-type SRVs in sample (contd.)}

\noindent Notes: $V$ amplitudes are usually given; in a few cases only $B$ or
photographic are available.

\noindent CSV 100168 has $M_K= -4.3$, NSV 03721 has $M_K= -4.11 $. 
These stars are below the cut-off of the figure.

\noindent GCVS: Combined General Catalog of Variable Stars (Samus et al, 2004), as
quoted on VizieR (CDS, Strasbourg).

\noindent P et al: Percy et al (2004); $^1$denotes first rank.  Amplitudes
of stars with P et al in the ref. column are in fact taken from PWH.

\noindent PDKT: Percy, Dunlop, Kassim \& Thompson, 2001. More certain
periods given in bold-face type; less certain periods denoted by a colon.

\noindent PNH: Percy, Nyssa \& Henry, 2001. Most secure periods given in
bold-face type; most uncertain periods are marked with a colon.

\noindent PWH: Percy, Wilson \& Henry, 2001. The periods marked with an
asterisk are the ones which seem to be most stable and well determined. The
amplitude of V450 Aql is taken from GCVS.

\noindent For L$_2$ Pup an average $V$ amplitude is given (Fig.\ 1, Bedding
et al, 2002). Its $K$ mag was taken to be --2.15 ({\it ibid}, footnote, p.\
81).

\label{sample}

\end{table*}

\section{The $M_K$, log$P$ diagram}

\begin{figure}
\epsfxsize=8.5cm
\epsffile[28 420 539 786]{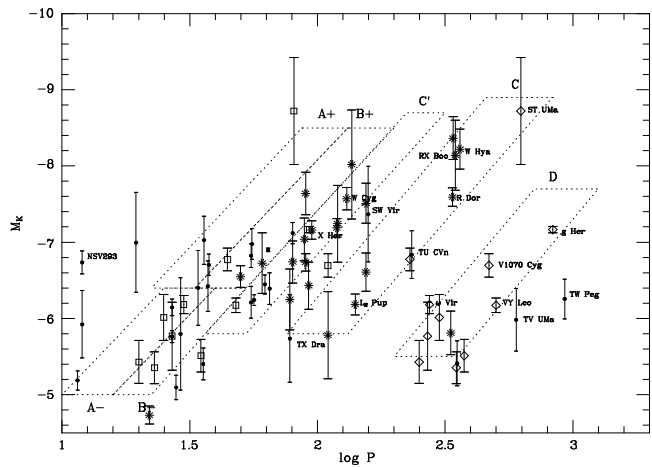}

\caption{$K$, log$P$ diagram for local M giants with parallaxes greater than 3
times the probable error. The positions of the LMC sequences defined by Ita
et al (2004) are shown as dotted boxes and labelled A+, B+ etc. Stars with
$V$ amplitude $>$ 1.0 are shown as asterisks. Double-period stars are shown
as boxes (short periods) and diamonds (long-periods).}

\label{klogp1}
\end{figure}

The $M_K$, log$P$ diagram can be presented in several different ways. Fig
\ref{klogp1} shows $M_K$ vs log$P$ for all the sources in the sample
with significant parallaxes ($\pi > 3\sigma_\pi$). The expected positions of
the various series as defined by Ita et al (2004) for the LMC are shown as
dotted parallelogram boxes, assuming a distance modulus of 18.5. The error
bars are based on the quoted probable errors in the parallax. Some of these
are quite large, leading to ambiguity as to which box a particular point
belongs to.

It can be seen that there are a few apparent exceptions to the Ita et al
classifications, such as the occurrence of single-period variables in or
near the D box and large-amplitude variables in the B$+$ box. Some of these
may simply be attributed to limited or poor data. Other effects are
discussed below in connection with Fig \ref{comparison}.

\begin{figure}
\epsfxsize=8.5cm
\epsffile[28 420 539 786]{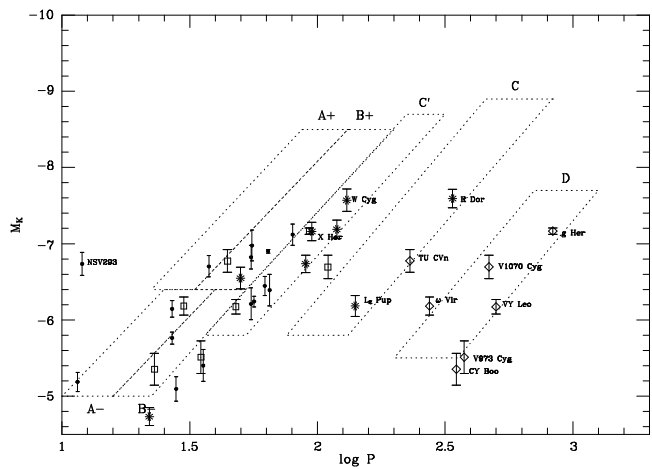}

\caption{$K$, log$P$ diagram for local M giants with parallaxes greater than
ten times the probable error. See Fig 1 for additional information.}

\label{klogp2}
\end{figure}

\begin{figure*}
\epsfxsize=15cm
\epsffile[28 403 539 786]{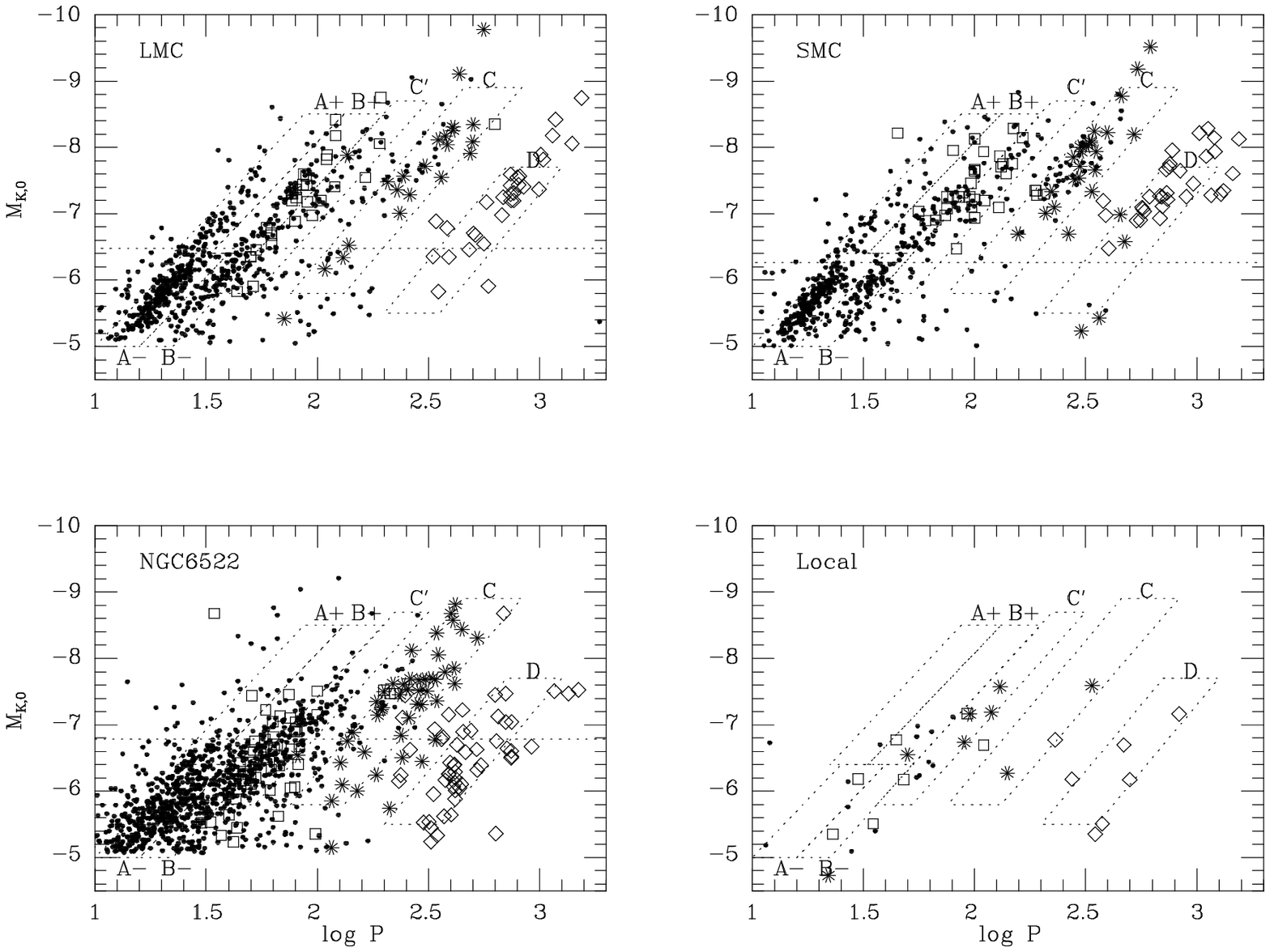}

\caption{Comparison of the $K$, log$P$ diagrams for SRVs in the two
Magellanic Clouds, the NGC\,6522 field in the Bulge and the solar
neighbourhood. The `Local' box is the same as Fig 2 except that the error
bars have been omitted. See Fig 1 for additional information. Notes: (a) the
distance moduli of the LMC, the SMC and the NGC\,6522 fields have been taken
to be 18.5, 18.94 and 14.7 respectively (b) the non-local $K$-band data have
been taken from the 2MASS survey (c) the scatter in the Bulge is
intrisically higher than in the other fields because of depth effects and
(d) the levels of the RGBs, indicated by dotted horizontal lines, are as
given in the text (e) Mira variables have not, in general, been included in
the Local sample.}

\label{comparison}
\end{figure*}

A more refined diagram is given in Fig \ref{klogp2}, where only sources with
parallaxes greater than ten times their probable error are shown. The
error bars are sufficiently small that most of the stars can be assigned to
the appropriate Ita et al (2004) boxes with confidence.

A general view of the $M_K$, log$P$ diagram for fields of different
metallicity is given in Fig \ref{comparison}. The data for the LMC, the SMC
and the NGC\,6522 Baade's Window field of the galactic bulge are taken from
Fig 5 of Schultheis, Glass \& Cioni (2004). The photometry of these fields
is from 2MASS and the periods are from MACHO, and have been derived in the
same way. Only the predominant periods have been included except that in
those cases with a long secondary period both have been plotted. Some stars
fall above the Ita et al boxes because the latters' data were taken with the
SIRIUS camera on the IRSF telescope at Sutherland and show saturation at
fainter limits. The levels of the tips of the RGBs have been taken to be
$M_{K,0}$ = --6.48 in the LMC and --6.26 in the SMC, based on the values
given by Kiss \& Bedding (2004). For the RGB tip in the NGC\,6522 field, the
difference of 0.3 mag in level from that of the LMC, determined by
Schultheis, Glass \& Cioni (2004), has simply been subtracted (to get
--6.78).

The amplitude criterion for large amplitude variables applied in the LMC,
SMC and NGC\,6522 is $\Delta r_{\rm MACHO} > 1.0$, whereas that for the
local sample is $\Delta V > 1.0$. Since the $V$ amplitudes are usually
somewhat greater than those at $r_{\rm MACHO}$, it is probable that more
local stars have been classified as having large amplitudes than should have
been. If $\Delta V > 1.63$ is adopted as the criterion, the only
large-amplitude stars are those in or close to box C.

There are several clear trends between the first three samples. The A$^+$,
B$^+$ and upper C$'$ sequences are relatively more populated in the
Magellanic Clouds than in the Bulge. The low ends of the C (Mira) and C$'$
sequences are relatively under-populated in the Magellanic Clouds.
Doubly periodic variables are sparse or lacking at the lower luminosity end
in the Magellanic Clouds. Further, the number of small amplitude variables
in the C box declines with increasing metallicity
(SMC$\rightarrow$LMC$\rightarrow$NGC\,6522).

As mentioned, Matsunaga et al (2006) have presented a preliminary $K$,
log$P$ diagram (their Fig 1) for galactic globular clusters (GCs). This may
be compared with Fig \ref{comparison}. No members of the A sequence have yet
been found among the GCs, though this may be a consequence of their smaller
amplitude and the fact that the survey was carried out in at $JHK'$. The
B$^+$ and C$'$ sequences in the GCs cut off at $M_K$ $\sim$ --6.5 and --6.3,
respectively, which is about the level of the RGB tip. The corresponding
level for the NGC\,6522 field is about --7.5. Even the C sequence Miras,
which are confined to the metal-rich clusters, reach no further than $M_K$
$\sim$ --7.6 in the GCs, as compared to $\sim$
--8.8 in the Bulge. Matsunaga et al attribute the effects they discuss to
differences in the Initial Mass Functions between the two fields (lower
masses in the GCs).

In comparing the local M-type SRVs with the others we must remember that
normal Miras have been omitted and that the relative lack of short-period
and small amplitude SRVs is largely a selection effect. Further, the most
luminous Magellanic Cloud SRVs are carbon stars, which are not present in
the Bulge and have not been included in the local sample.

The more luminous part of the local M-star sample should be directly
comparable to the Bulge field, where there are no C stars. The A$^+$, B$^+$
and C$'$ sequences seem to cut off at the same levels, both of which are
well below the luminosities reached in the Magellanic Clouds. This is also
true of the D sequence, though the latter is not a truly independent one,
being composed of stars that have simultaneous shorter periods in the B and
C$'$ sequences. The lower luminosity ends of the C and D sequences can also
be compared, since these stars have moderately large amplitudes and hence
are easier to detect as variables. Again, though the significance of the
conclusion is limited by the small numbers, the parameter space occupied by
the members of each field is similar.

The conclusion seems to be that the local M giant SRVs compare most closely
with those in the NGC\,6522 field.

\section{Prevalence of mass loss}

Early results from the ISOGAL survey at 7 and 15$\mu$m using the ISOCAM
camera of the ISO satellite showed from the [15], [7] -- [15] diagram that
there is a sequence of increasing mass-loss from early to late-type M stars
on the AGB in the Bulge (Glass et al,1999; Omont et al, 1999). It is not
confined to the known Mira variables; substantial mass-loss also occurs in
other late-type M giants. See Glass \& Schultheis (2002) for an analysis of
the complete sample of M giants observed by Blanco (1986) in the NGC\,6522
Baade's Window clear field towards the galactic bulge.
 
The presence of mass-loss from the present sample of SRVs is seen from their
$K$ -- [12] colours.  The $K$-band is contributed to mainly by the stellar
photospheres of these stars but the [12] band, as the ISO 15$\mu$m band, can
be heavily affected by radiation from dust in circumstellar shells. The [12]
mags for almost all the stars in Table \ref{sample} are available from the
IRAS Point Source Catalog (Beichman et al, 1988).  Miras have much smaller
amplitudes in the infrared than in the visible and, by analogy, the
amplitudes of SRVs in the infrared are expected to be modest, of order a few
tenths of a mag at maximum. For example, the short-term variability of L$_2$
Pup at $K$ is about a quarter of that at $V$ (Bedding et al, 2002).
Nevertheless, some degree of uncertainty in their $K$ -- [12] colours must
be expected from variability. The calibration of the relation between \.{M}
($M_{\odot}$ yr$^{-1}$) and $K$ -- [12] has been discussed by Whitelock et
al (1994); saturation is reached at $K$ -- [12] $\sim$ 5, corresponding to a
mass-loss rate of about log \.{M} $\sim$ --5. The dust mass-loss rates from
SRVs in the LMC, the SMC and the NGC\,6522 fields appear to be fairly
similar from ISO data (Schultheis, Glass \& Cioni, 2004, Fig.\ 16).

\begin{figure}
\epsfxsize=8.5cm
\epsffile[28 431 539 786]{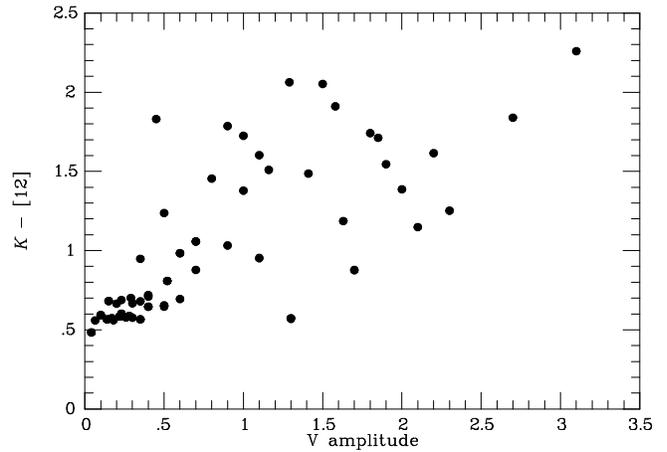}

\caption{$K$ -- [12] colour vs visual amplitude for the SRVs. The minimum
value of $K$ -- [12] $\sim$ 0.5 is appropriate to early M-type giant
photospheres the values are higher when dust shells are present. While there
is not a one-to-one correlation, high visual amplitude appears to be a
pre-requisite for dust emission.}

\label{XSvsA}
\end{figure}

Fig \ref{XSvsA} shows that infrared excesses are associated with large
amplitude of variation. The minimum of $K$ -- [12] colour for the sample is
offset by about 0.5 mag, which is appropriate for early M-type giant
photospheres. It is based on a 12 $\mu$m mag of 0 corresponding to 28.3 Jy
(Beichman et al, 1988).

\begin{figure}
\epsfxsize=8.5cm
\epsffile[28 420 539 786]{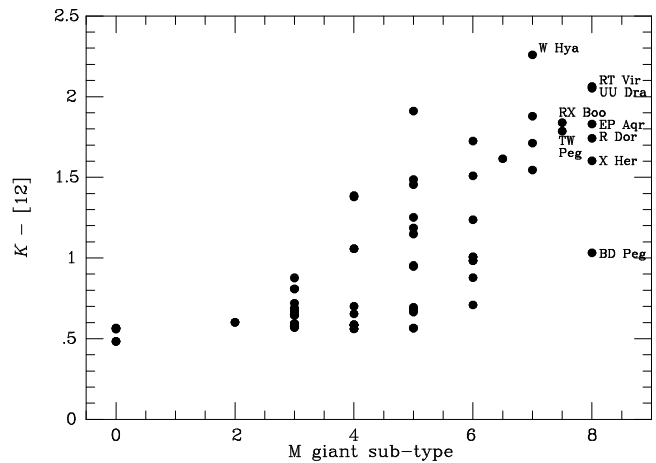}

\caption{$K$ -- [12] colour vs M giant sub-type. Mass-loss increases towards
later sub-types, as seen also in the NGC\,6522 field (Glass \& Schultheis
(2002).}

\label{CvsM}
\end{figure}

The $K$ -- [12] colour is also shown as a function of M giant sub-type in Fig
\ref{CvsM}. Colour excesses begin to appear at M3, 
as also found by Glass \& Schultheis (2002) in an analyis of a complete
sample of M giants observed by Blanco (1986) in the NGC\,6522 Baade's Window
clear field towards the galactic bulge.  

\begin{figure}
\epsfxsize=8.5cm
\epsffile[28 420 539 786]{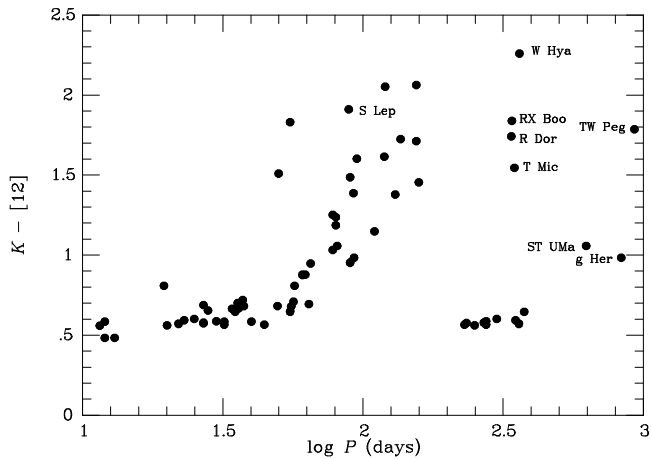}

\caption{$K$ - [12] colour, indicative of excesses arising from dust shells,
vs log(period). Significant mass-loss starts at periods of 60--70 days. The
Mira-like variables W Hya, RX Boo and R Dor stand out at long periods. Many
of the long periods correspond to double-period stars and show no evidence
for shells; however ST UMa and g Her are from this category and
exceptionally have excesses.}

\label{CvsP}
\end{figure}

Fig \ref{CvsP} shows the $K$ -- [12] colour vs log (period) for the sample.
There is a very clear increase of infrared excess with period, starting at
about log $P$ = 1.75 ($P$ = 56 days). This is in agreement with the results
of Alard et al (2001) for the NGC\,6522 Baade's Window variables as well as
the LMC and SMC fields of Glass, Schultheis and Cioni ((2004).

Of the 64 variables listed in Table 2, 35 have had their IRAS (10$\mu$m
region) spectra classified by Sloan \& Price (1998). Ten of them show
`naked' photospheres; i.e., dust shells were not detected. They correspond
to the shortest-period SRVs. The remainder are classified as types SE1 to
SE8, according to the strength of their SiO features. Many are also
classified as `t', meaning that they show the 13$\mu$m feature. The SE
sub-classes and presence or absence of the `t' show no correlation with $K$
-- [12] colour.

Given the fairly clear dependence of $K$ -- [12] colour on M-giant sub-type
and log $P$ seen in Figs \ref{CvsM} and \ref{CvsP}, it is surprising that
Olofsson et al (2002), using mass-loss rates for SRVs based on CO radio
data, do not see clear correlations with pulsation period or stellar
blackbody temperature (their Figs 7 and 8).

\section{Prospects for progress}

The current picture of SRVs in the solar neighbourhood, though improved in
detail, remains sketchy because of the small size and haphazard nature of
the sample. Enough evidence now exists, however, to show that local SRVs
occupy the same areas of the $K$, log $P$ diagram as stars in the NGC\,6522
field of the Bulge and that they obey similar $K$ -- log $P$ relations.

The currently available data, especially at the short-period end, are too
sparse. Hipparcos parallaxes are available for numerous early M-type
stars which have not yet been monitored with sufficient accuracy or for long
enough times to find their variability properties. While these stars are
usually too bright to be included in current all-sky monitoring projects,
they are suitable for photometric measurements with small telescopes. Thus,
frequent measurements over periods of one year or more, though tedious, can
certainly be contemplated.

Because the Hipparcos parallaxes are directly the result of trigonometrical
determinations and often of high accuracy, an increase of the number in the
sample will yield a sounder absolute calibration of the properties of the
SRVs and other M-type giants. With a better understanding of the effects of
metallicity and age, they may even prove useful as distance indicators. 

Finally, we note that $K$-band observations for bright stars are fortunately
still possible.

\section{Acknowledgments}

An earlier version of this paper, based on the published Hipparcos
parallaxes, was presented at the conference ``Why Galaxies Care About AGB
Stars", Vienna, August 6--11, 2006.

We wish to acknowledge use of data from the Centre des Donn\'{e}es
Stellaires, Strasbourg.  Drs T. Lloyd Evans, M. Schultheis and A. Zijlstra
read drafts of the paper, providing useful comments and some extra data.

\section{Appendix}

\end{document}